\documentclass[10pt, letterpaper]{IEEEtran}
\makeatletter
\def\ps@headings{%
\def\@oddhead{\mbox{}\scriptsize\rightmark \hfil \thepage}%
\def\@evenhead{\scriptsize\thepage \hfil \leftmark\mbox{}}%
\def\@oddfoot{}%
\def\@evenfoot{}}
\makeatother
\usepackage{amssymb,amsmath,amsthm}
\usepackage{graphicx}
\usepackage{caption}
\usepackage{esvect}
\usepackage{changes}
 \newcommand{\rem}[1]{}

\usepackage[TABBOTCAP]{subfigure}
\usepackage{mdwmath}
\usepackage{mdwtab}
\usepackage{soul}
\usepackage{cite}
 \usepackage{multirow}
 \usepackage{color}
  \usepackage[linesnumbered,ruled]{algorithm2e}
\usepackage{setspace}

\begin{document}
    \title{Computationally Secure Optical Transmission Systems with Optical Encryption at Line Rate}
	\author{Anna Engelmann and Admela Jukan}
   \maketitle

\begin{abstract}
We propose a novel system for optical encryption based on an optical XOR and optical Linear Feedback Shift Register (oLFSRs). Though we choose LFSR for its ability to process optical signals at line rate, we consider the fact that it offers no cryptographic security. To address the security shortfall, we propose implementation of parallel oLFSRs, whereby the resulting key-stream at line rate is controlled electronically by a nonlinear random number generator at speeds much lower than the optical line rate, which makes the system practically relevant. The analysis of computational security shows that the proposed system is secure against wiretapping and can be engineered with the state of the art optical components. 
 \end{abstract}

\section{Introduction}

Making all-optical encryption practical today would require two basic components: an optical encryption component and an optical key generator. For their realization, not only these components need to be developed but also significant system design obstacles need to be overcome, whereby the main challenge is encryption and key generation at the optical channel line rate. It is therefore that in today's systems, the payload of an OTN container is encrypted in the electronic layer and the so-called Advanced Encryption Standard (AES) does not perform encryption and decryption at line speed \cite{1}. 

\par Currently, two state-of-the-art optical components can help implement all-optical encryption system at line speed: optical XOR and optical Linear Feedback Shift Register (oLFSRs). Optical XOR has been proposed for encryption, as it can easily concatenate plaintext with a key \cite{4,6,16,3}. For optical key generation, no solution currently exist. Our idea to use an oLFSR for key generation at line speed falls short, on the other hand, since the LFSR-based ciphers are known for their weak cryptographic security due to the linear properties \cite{9}. Since LFSR is simple to implement all-optically based on the concatenated optical XORs and a shift register \cite{20, 16}, we here address the fundamental security shortfall. 

We propose a new and practical solution towards optical encryption and key generation, based on \emph{parallel} oLFSRs, which are controlled and periodically \emph{reseeded} by a pseudo random number generator (RNG) at speeds much lower than the rate of optical channel, which makes the system practically relevant. Under these assumptions and presence of guessing wiretapper in optical network, we analyze the computational security of the proposed system and derive important system parameters, including  the minimal polynomial degree and the number of different polynomials required for oLFSR implementation, the optimal key length generated by oLFSR, and the implementation overhead. The results show that the proposed optical encryption system provides high level of computational security and thus carries promise for further studies. 


\section{System Model}
\begin{figure*}[!t]
\centering
\includegraphics[width=0.93\textwidth]{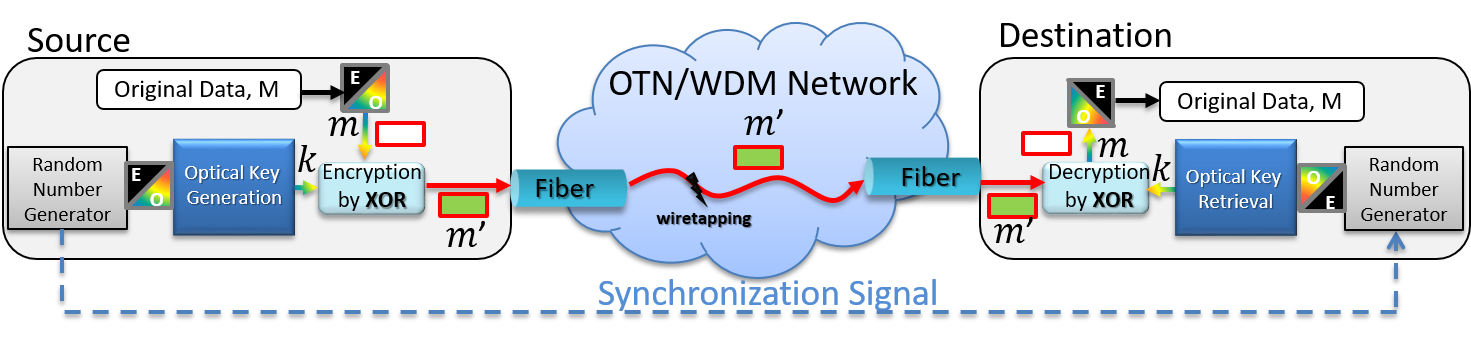}%
  \vspace{-0.1cm}
  \caption{All-optical transmission system with encryption capability. $k$: generated key of length $L_k$; $m$: secret message of length $L_m$; $m'$: encrypted message of length $L_{m'}$; $M$: original optical data of length $L_M=N\cdot L_m$}
 \vspace{-0.4cm}
  \label{net}
\end{figure*}

\par Fig.\ref{net} illustrates the optical transmission system envisioned.  As the optical network usually encompass both the data plane, and from the data plane a separated control plane, we assume that the control plane is able to monitor information about network topology, assign wavelengths, and control the configuration of the encryption and key generation components. However, the control plane does not distribute the actual session keys, but only the synchronization signal for the Pseudo Random Number Generator (RNG) for random selection of parameters for session key generation and interleaving of original data in end-systems prior to data encryption. The bit interleaving can be defined and configured by control plane for each end-to-end connection and is required to preliminary hide the content of optical data as well as to force an attacker to decrypt a whole wiretap data for content recovery.

\par In the source, the original optical bit stream $M$ is interleaved and virtually split into multiple optical units $m$. Each optical data $m$ is encrypted with the key $k$ from the Optical Key Generator (OKG). This is done by applying optical XOR. The encrypted $m'$ is sent to the optical transmission channel provisioned. When any encrypted parts $m'$ of $M$ reaches the destination, it is decrypted with key $k$ and converted into electronic signal for following deinterleaving, i.e., recovering of original data. Due to the fact that RNG in source and destination are synchronized, both generate the same pseudo random numbers, i.e., generator polynomials $c^*$ and seeds $h$ to define and initialize LFSR, respectively. As a result, OKG generates the same keys $k$ for en- und decryption.

\par We propose to use all-optical XOR gate \cite{3,5,8} for encryption. From the security perspective, the XOR operation transforms the original data into encrypted data and, thus, can prevent wiretapping, whereby each incoming plaintext $m$ of length $L_m$ in source is mixed, i.e., XOR concatenated, with a session key $k$ of length $L_k$ into encrypted data $m'=[m\oplus k]$ of the length $L_{m'}=L_m$. The encrypted data $m'$ is decrypted at the destination as $m'\oplus k=m\oplus k\oplus k=m$. Here, the ultrafast nonlinear interferometers based on semiconductor optical amplifiers (SOAs) can be used to combine two optical streams at line rate up to $160 $Gb/s \cite{3}, whereby transverse electric (TE) and transverse magnetic (TM) components of a probe pulse can be split and recombine by setting the relative optical delays between them. The two data modulated pulsed signals to undergo XOR operation, secret $m$ and key $k$, are assumed to have equal bit rate up to $160 $Gb/s.

\subsection{Key generation with oLFSR}
\par
\begin{figure}[t]
 \centering
\includegraphics[width= 0.475\textwidth]{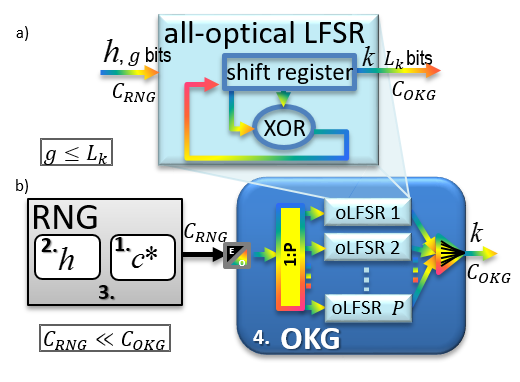}
 \vspace{-0.1cm}
\caption{Key generation with oLFSRs. $c^*$: bit stream to define primary oLFSR; $h$: seed as a bit stream of length $g$;}
  \vspace{-0.4cm}
\label{ParLFSR}
\end{figure}

 LFSR has been traditionally used for random number generation \cite{17,9} and not for key generation, due to the known weak cryptographic security. In its basic configuration, as illustrated in Fig. \ref{ParLFSR}a based on \cite{16}, the shift register has a fixed size of $g$ bits, whereby the seed $h$ of $g$ bits presents the initial sequence. From this register, a set of fixed bits denoted as $i$ and $j$, corresponding to the generator polynomial $c^*$, are XOR concatenated and the resulting bit is fed to the shift register at the last position ($g$). After that, the sequence in the register is 1-bit shifted. To address the issue of weak security, we use a nonlinear pseudo RNG at comparably lower line rate in combination with  parallel LFSRs, and deploy reseeding of LFSR during key generation. With a proper choice of system parameters, we later show that we are able to implement a secure key generation. 

\par Fig. \ref{ParLFSR}b illustrates the key generation. To generate a session key $k$, OKG uses a generator polynomial $c^*$ of degree $g$ and seed $h$, both randomly selected by RNG. Since RNGs in source and destination are synchronized, i.e., able to generate the same polynomials $c^*$ and seeds $h$, the key generated can be \textit{infinitely long}.  We propose to utilize $P$ generator polynomials to assign the same number of corresponding parallel oLFSR, whereby each polynomial corresponds to one oLFSR configuration. In our system, a randomly selected seed $h$ is forwarded towards its corresponding oLFSR as selected by an optical $1:P$ switch. The resulting optical signal from that oLFSR is finally forwarded as an encryption key $k$ to the optical XOR. It should be noted that the RNG provides a \emph{true secret key} stream at lower line rates, i.e., electronic bit streams at rate $C_{RNG}\leq 1.3$ Gbit/s \cite{18,15} to define generator polynomial (oLFSR) and seeds. The oLFSR allows us to extend this secret to a comparably longer optical encryption key $k$, namely at line rates of $C_{OKG}\leq 250$ Gbit/s \cite{6}. Thus, oLFSR makes it possible to adapt the length and bit rate of \emph{true secret key} at the optical line rate, which is the system salient feature. 

\par In summary, the system proposed addresses the issue of weak cryptographic security of the LFSR through the following simple system modifications, including use of: (1) different generator polynomials $c^*$ of any degrees $g$, and (2) randomly selected seeds $h$ to start off the LFSR; (3) reseeding, i.e., periodical selecting of new seeds during generation of certain key with defined polynomial, and (4) implementation of reconfigurable OKG with parallelized LFSRs. The reseeding can help us to increase the amount and randomness of bit sequence generated by RNG, whereby RNG can generate new seed continuously or periodically, depending on configuration; for instance, the triggering can be configured by the control plane to occur periodically. It should be noted that using different polynomials with optical components is a challenge. As already studied elsewhere, in a system with different polynomials implemented in electronics, each polynomial would have to be implemented as a separate LFSR \cite{6} or by extending of logic gates \cite{17}, which does not scale. The idea of parallel oLFSR and optical switching thus replaces the need for periodical polynomial reconfiguration, which can today only be implemented in electronics. While the chosen oLFSR generates a part of an infinite long session key, RNG can then define the next random seed for its initialization, to be cyclically reseeded, etc.

\subsection{Attacker Model}
\par We analyze the wiretapping attack, whereby the attacker is always able to access the fiber link and, thus, any encrypted part $m'$. Due to bit interleaving of data $M$ before optical encryption, the secret content can be only revealed, if a whole optical data $M$ is decrypted and deinterleaved. Thus, the attacker must guess any secret keys $k$ first. In case an attacker can access the end system and analyze the generation technique of session key, the probability for breaking the LFSR at the first try is $100\%$: an attacker is able to break LFSR based on generator polynomial of degree $g$ with a time complexity defined as $O(g^2))$ and  memory complexity $O(2g)$, if receiving $2g$ bits generated by LFSR. Considering this weakness, we assume that the attacker already knows the  structure of OKG based on LFSRs. All that is left then to guess a correct generator polynomials and correct sequence of seeds. 





\section{System analysis}

\par This section focuses on  computational security, i.e., complexity to find out the right generator polynomial and the right sequence of seeds utilized to generate $k$. We analyze the required number of parallel LFSRs $P$ and number of required reseeds $N$ to make the system computationally secure. We also discuss bounds of the parameters required for OKG implementation. 

\subsection{Security analysis}

For the analysis, we assume that all primitive irreducible polynomials of any degree are known and can be utilized for initialization of oLFSR. We assume that RNG can randomly select one out of $P\leq\varphi(2^g-1)/g$ generator polynomials of any degree $g$, $g_{min}\leq g\leq g_{max}$, where $\varphi(x)$ is Euler function. Generally, the minimal degree $g_{min}$ should be defined so that the key length generated by LFSR is larger than the length of secret data $m$ encrypted by this key $k$, i.e., $L_m\leq L_k<2^{g_{min}}-1$ and $g_{min}\geq log_2(L_k+1)$, to avoid the cyclic repeat of a key. However, due to the fact that the length of optical data $M$ can be very large, i.e., multiples of $L_m$, $L_M=N\cdot L_m$, we can relax the condition $g_{min}\geq log_2(L_M+1)$ for minimal polynomial degree $g_{min}$ as follows $g_{min}\geq log_2(\frac{L_M}{N}+1)$, where $N$ describes the number of reseeds.

When utilizing polynomials of any degree $g$, the parameters $c^*$ and $h$ of length $g$ are chosen randomly, and kept secret. Thus, the entropy of true secret key can be defined as $H_1 (k_{RNG})=log(P (2^g-1))$, while one out of $P$ existing generator polynomials of degree $g$ and a seed out of $(2^g-1)$ can be selected for generation of encryption key $k$. From the attacker's perspective, the session key $k$ can be an arbitrary bit sequence out of $2^{L_{m'}}$ possible ones, i.e., the entropy can be defined as $H_2 (k)=L_{m'}$ bits. Generally, an attacker can follow the algorithm for generation of key $k$ and, thus, either guess any $c^*$ and $h$, or directly guess a bit sequence $k$ of length $L_{m'}$. In the former case, the equivocation is defined as $H_1(k|m')=log_2(P (2^g-1))=H_1 (k_{RNG})$. In the latter case, $H_2(k|m' )=L_{m'}=H_2 (k)$. It is faster however to guess the parameters $c^*$ and $h$, if $P (2^g-1)<2^{L_{m'}}$. Thus, the equivocation $H_1 (k|m')$ must be equal to or larger than entropy $H_2 (k)$ for a perfect secrecy \cite{10}. The resulting condition for a perfect secrecy can be defined as
\begin{equation}\label{optLen}
log_2\left(P(2^g-1)\right)\geq L_k
\vspace{-0.25cm}
\end{equation} 
Eq. \eqref{optLen} also defines the maximal size of optical data $L_m$ encrypted before the new cycle of oLFSR reseeding. For a strong practical security, the system must be unbreakable also in case of a brute force attacks (BFA), whereby the key computation complexity, i.e., time overhead, must be high. Since an attacker must check all combinations of polynomials and seeds, the time required to break the proposed system is 
\begin{equation}\label{T1}
T^{BFA}=N\cdot \tau\cdot P\cdot(2^g-1),
\vspace{-0.25cm}
\end{equation}
where $\tau$ is a time required for decryption with one key guessed by the attacker, while key guessing time (generation of one binary sequence $k$ with known LFSR) is assumed as zero, and $N$ is a number of reseeds.
Since secret optical data flow $M$ is, generally, larger than the optimal key length defined by Eg.~\eqref{optLen}, i.e., $L_M >L_k=L_m$, the oLFSR must be either reseeded every time $L_m$ bits of optical data are encrypted. Thus, the number of key reseeds $N$ depends on optimal key length $L_k$ and data flow length $L_M $ and can be defined as
\begin{equation}\label{resets}
N=L_M/L_k
\vspace{-0.25cm}
\end{equation}
When generator polynomials utilized in our system are predefined and oLFSR could be generated on the fly (which would only be possible in electronics), they also need to be stored. The buffer size required is a function of degree $g$:
\begin{equation}\label{buffer}
S=P (g+1)
\end{equation}

\subsection{Discussion on implementation}
The utilization of all polynomials of given degree $g$, are very hard to implement in practice and especially all-optically. Thus, we propose to bound $P<<\varphi(2^g-1)/g$ primitive irreducible polynomials of the same degree $g$ to correspond to all-optical LFSRs and use them in parallel, whereby only one oLFSR is  selected and periodically reinitialized by seed to generate a part $L_k$ of a session key. Since there is a need for erasing of shift register (initialization) and for skipping the first output bits related to random seed, oLFSR initialization can lead to key generation interruptions and decrease in bit rate of OKG. Thus, there is a need for optical buffer and controlled signal clocking. However, we envision, that the oLFSR initialization will be most efficient, when another oLFSR based on the same generator polynomial is generating a key, whereby the duration of key generation period must be at least the same as duration of oLFSR initialization.  To this end, we propose to relax the condition for optimal key length defined by Eq. \eqref{optLen}, as $L_k=n\cdot g$, $n\geq 1$, and analyze next the practical security of proposed reseeding method.

Let us now reverse engineer the number of oLFSRs and its length required given the time needed for a possible BFA, where $T^{BFA}$ is set to a very large number of years and the time for one decoding try $\tau$ is defined by the state of the art technology and known. When the OKG consists only of $P$ oLFSRs (each implementing one polynomial of degree $g$), the Eq. \eqref{T1} can be modified as $T^{BFA}\leq N\cdot \tau\ P(2^g-1)$, where the number of required reconfigurations is defined by Eq.\eqref{resets} and with $L_k=n\cdot g$ modified as $N=L_M/n\cdot g$. As a result, boundary condition for strong practical security is defined as
\begin{equation}\label{Cond}
\frac{n\cdot T^{BFA}}{\tau L_MP}\leq \frac{2^g-1}{g},
\end{equation}
where the number of parallel oLFSRs $P$ is limited as $2\leq P\leq \varphi(2^g-1)/g$ and $1\leq n\leq \tfrac{2^g-1}{g}$.

\section{Numerical Results}
\par In this section, we show the numerical results for the previous analysis. First, we consider all possible generator polynomials of degrees from $10$ to $45$, i.e., $P=\sum_{g=10}^{45}\varphi(2^g-1)/g$. To define a range of required generator polynomials, i.e., $c^*$, we focus on two case studies: C1) $g_{min}=10$ is fixed and $g_{max}$ is variable; C2) $g_{max}=45$ is fixed and $g_{min}$ is variable. We assumed that optical bit flow $M$ has a size $L_M=1.25$ Gbits (e.g., an OTN transmission unit) and decryption time $\tau$=$10^{-18}$ sec based on data from Aurora \cite{11} at $180$ Petaflops.

\par
\begin{figure}[t]
 \centering
\includegraphics[width= 0.47\textwidth]{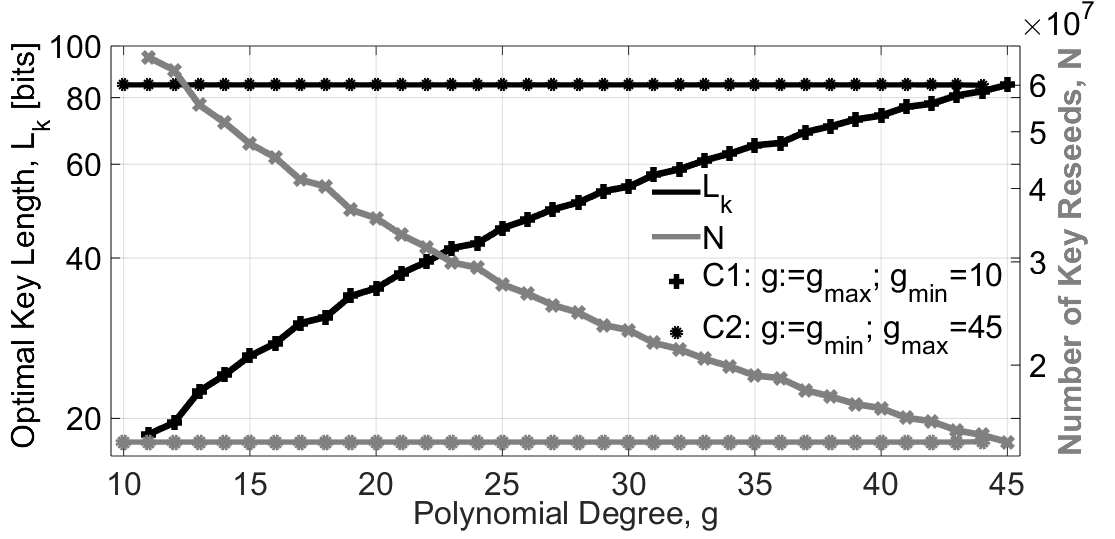}
    \vspace{-0.2cm}
  \caption{Optimal key length $L_k$ and nr. of resets $N$ vs. $g$.}
\label{fig1}
\end{figure}

 Fig. \ref{fig1} shows optimal key length $L_k$ defined by Eq.\eqref{optLen} and the required number of key reseeds $N$ as a function of polynomial degree $g$. For C2, $L_k$ is constant (around $84$ bits) for any $g:=g_{min}$. For C1, the optimal key length $L_k$ increases with increasing $g_{max}$. However, the key length directly defines the number of reseeds. In contrast to C2, where $N$ is minimal and constant, the number of reseeds, in case of C1, decreases with increasing $g_{max}$. 

\par
\begin{figure}[t]
 \centering
\includegraphics[width= 0.47\textwidth]{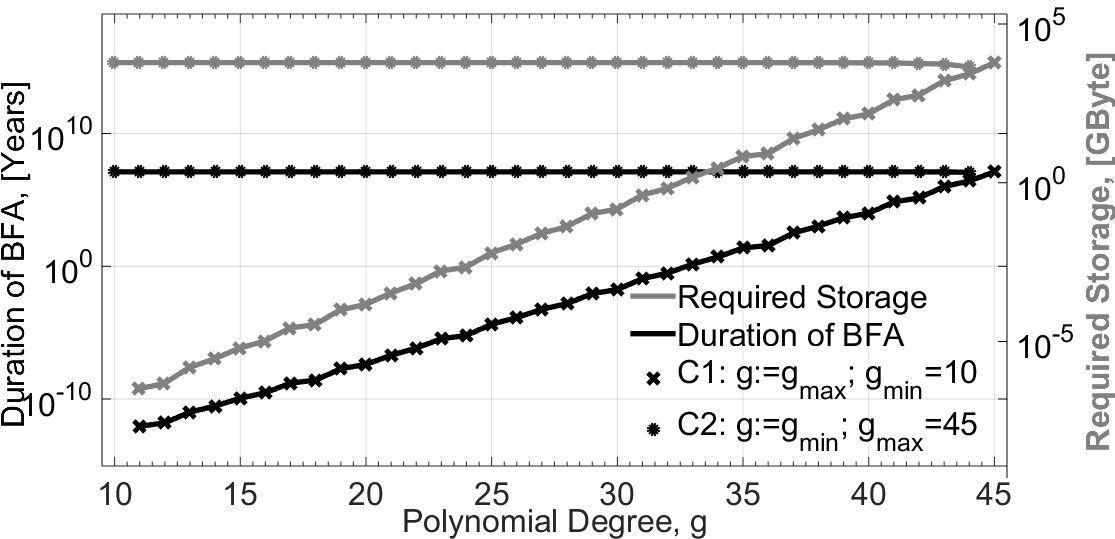}
    \vspace{-0.2cm}
  \caption{Duration of BFA and required storage vs. $g$.}
\label{fig3}
\end{figure}

 Fig.~\ref{fig3} shows the time (in years) required to decode a wiretapped optical data of size $L_M$ in case of Brute Force Attack (BFA) calculated with Eq. \eqref{T1} as well as the storage required in the node. As it can be seen, a prohibitively long time of over $200$ years can be measured in case of $C1$ for $g=37$. In this case, around $24$ GByte additional storage is required to store the generator polynomials. In C2, an attacker requires more than $1.8\cdot10^{8}$ years to decrypt a whole data flow, whereby this time is constant for all $g_{min}=g$. On the other hand, it requires prohibitively large storage of $10^{4}$ Gbyte. In general, larger the polynomial degree $g$ better the security.

\par
\begin{figure}[t]
 \centering
\includegraphics[width= 0.47\textwidth]{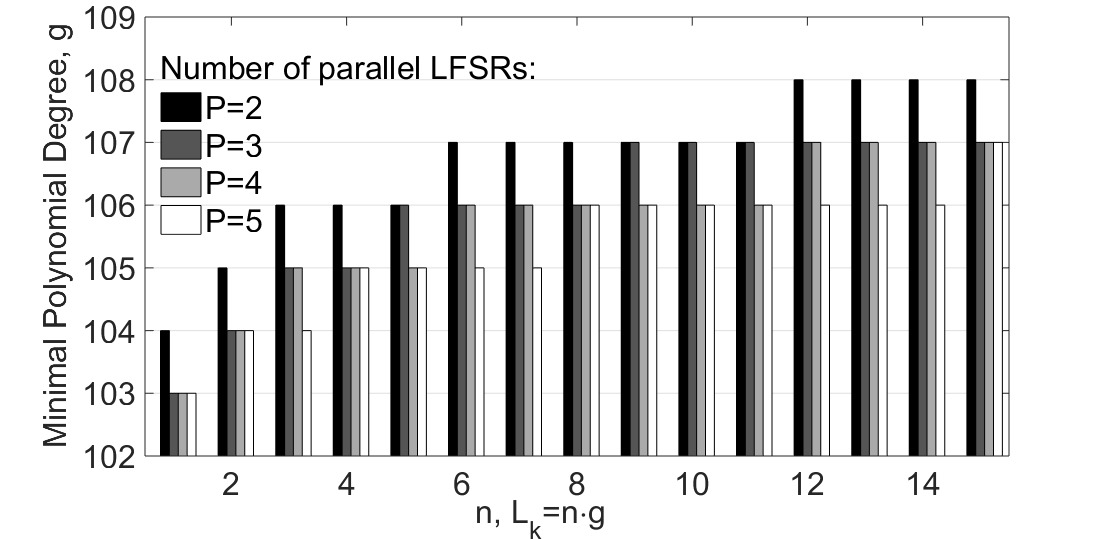}
  \vspace{-0.2cm}
  \caption{Polynomial degree vs. key length vs. nr. of oLFSRs.}
\label{fig4}
\end{figure}

 Fig.~\ref{fig4} shows the boundary conditions for practical realization of all-optical key generator based on a few generator polynomials. It shows the minimal required  polynomial degree as a function of amount of parallel oLFSRs and the key length generated prior to the reseed. Here, we assumed the duration of BFA as $10^{13}$ years, as in case of BFA on AES key of length $128$ bits (calculated under assumptions made). The increasing length of key generated $L_k=ng$ with the same seed increases the required polynomial degree $g$. However, the increasing number of parallel oLFSRs utilized decreases the polynomial degree required. For example, OKG based on only $3$ oLFSR (g=106) can provide the same time complexity for BFA as OKG based on $2$ oLFSRs of the same length, if the maximal key length $L_k$ generated between reseeds is bounded as $8g$ and $5g$, respectively.  

\section{Conclusion}
We proposed for the first time to utilize all-optical XOR technique for encryption in the optical layer at line rate, with a technique based on  parallel all-optical LFSRs for generation of infinite long key. We showed that using different generator polynomials of defined degree, which are periodically reseeded, can provide a high practical security of optical data transmitted. The main results of this study is that the all-optical implementation of the key generator which is based on only $3$ optical LFSRs of an optimized length of $106$ bits provides high computational security (BFA takes $10^{13}$ years), whereby $848$ bits of the key can be generated without reseeding. 

\ifCLASSOPTIONcaptionsoff
  \newpage
\fi


 \bibliographystyle{IEEEtran}
\bibliography{bibL}

\begin{thebibliography}{10}
\providecommand{\url}[1]{#1}
\csname url@samestyle\endcsname
\providecommand{\newblock}{\relax}
\providecommand{\bibinfo}[2]{#2}
\providecommand{\BIBentrySTDinterwordspacing}{\spaceskip=0pt\relax}
\providecommand{\BIBentryALTinterwordstretchfactor}{4}
\providecommand{\BIBentryALTinterwordspacing}{\spaceskip=\fontdimen2\font plus
\BIBentryALTinterwordstretchfactor\fontdimen3\font minus
  \fontdimen4\font\relax}
\providecommand{\BIBforeignlanguage}[2]{{%
\expandafter\ifx\csname l@#1\endcsname\relax
\typeout{** WARNING: IEEEtran.bst: No hyphenation pattern has been}%
\typeout{** loaded for the language `#1'. Using the pattern for}%
\typeout{** the default language instead.}%
\else
\language=\csname l@#1\endcsname
\fi
#2}}
\providecommand{\BIBdecl}{\relax}
\BIBdecl

\bibitem{1}
\BIBentryALTinterwordspacing
``Fsp 3000,'' 2014. [Online]. Available:
  \url{http://www.advaoptical.com/home/products/
  scalable-optical-transport/fsp-3000}
\BIBentrySTDinterwordspacing

\bibitem{4}
X.~Y. et~al., ``Simple 40 gbit/s all-optical xor gate,'' \emph{Electronics
  Letters}, vol.~46, no.~3, pp. 229--230, 2010.

\bibitem{6}
X.~Z. et~al., ``High-speed all-optical encryption and decryption based on
  two-photon absorption in semiconductor optical amplifiers,'' \emph{IEEE/OSA
  Journal of Optical Communications and Networking}, vol.~7, no.~4, pp.
  276--285, 2015.

\bibitem{16}
M.~S. et~al., ``Optical linear feedback shift register,'' in \emph{CLEO
  EUROPE/EQEC}, 2011, pp. 1--1.

\bibitem{3}
E.~D. et~al., ``All-optical xor gate using single quantum-dot soa and optical
  filter,'' \emph{Journal of Lightwave Technology}, vol.~31, no.~23, pp.
  3813--3821, 2013.

\bibitem{9}
K.~Z. et~al., ``Pseudorandom bit generators in stream-cipher cryptography,''
  \emph{Computer}, vol.~24, no.~2, pp. 8--17, 1991.

\bibitem{20}
Y.~M.~J. et~al., ``All-optical circular shift register using semiconductor
  optical amplifiers,'' in \emph{International Conference on Photonics in
  Switching}, 2006, pp. 1--2.

\bibitem{5}
Y.~T. et~al., ``Simulation and demonstration of directed xor/xnor logic gates
  using two cascaded microring resonators,'' \emph{IEEE Photonics Journal},
  vol.~8, no.~2, pp. 1--11, 2016.

\bibitem{8}
X.~T. et~al., ``Experimental demonstration of high-speed logic gates of or,
  and, xor and nor in optical domain based on a single i/q modulator and direct
  detection,'' in \emph{COMCAS}, 2015, pp. 1--3.

\bibitem{17}
J.~M.-S. et~al., ``Multiple-polynomial lfsr based pseudorandom number generator
  for epc gen2 rfid tags,'' in \emph{IECON 2011}, 2011, pp. 3820--3825.

\bibitem{18}
M.~A. et~al., ``Design of a pseudo-chaotic number generator as a random number
  generator,'' in \emph{COMM}, 2016, pp. 401--404.

\bibitem{15}
A.~M. et~al., ``High speed and secure variable probability pseudo/true random
  number generator using fpga,'' in \emph{SIITME}, 2015, pp. 323--328.

\bibitem{10}
C.~E. Shannon, ``Communication theory of secrecy systems,'' \emph{The Bell
  System Technical Journal}, vol.~28, no.~4, pp. 656--715, 1949.

\bibitem{11}
\BIBentryALTinterwordspacing
``Aurora,'' 2016. [Online]. Available: \url{http://aurora.alcf.anl.gov/}
\BIBentrySTDinterwordspacing

\end{thebibliography}
\end{document}